\documentclass[12pt,preprint]{aastex}

\usepackage{multirow}
\usepackage[latin1]{inputenc}

\newcommand{\Bag}{\ensuremath{\mathsf{B}}}
\newcommand{\unit}[1]{\ensuremath{\, \mathrm{#1}}}

\shorttitle{Magnetars as Highly Magnetized Quark Stars}
\shortauthors{Orsaria et al.}

\begin{document}


\title{MAGNETARS AS HIGHLY MAGNETIZED QUARK STARS:\\
AN ANALYTICAL TREATMENT }

\author{M. Orsaria\altaffilmark{1,2} Ignacio F. Ranea-Sandoval\altaffilmark{1,2} and H. Vucetich\altaffilmark{1}}
\affil{$^1$ Gravitation, Astrophysics and Cosmology Group \\ Facultad de Ciencias Astron{\'o}micas y Geofísicas,\\
Paseo del Bosque S/N (1900),\\
 Universidad Nacional de La Plata UNLP, La Plata, Argentina\\
 $^2$ CONICET, Rivadavia 1917, 1033 Buenos Aires, Argentina}


\begin{abstract}
We present an analytical model of a magnetar as a high density magnetized quark bag.
The effect of strong magnetic fields ($B > 5 \times 10^{16}\unit{G}$) in the equation
of state is considered. An analytic expression for the mass-radius
relationship is found from the energy variational principle in
general relativity. Our results are compared with observational
evidences of possible quark and/or hybrid stars.
\end{abstract}

\keywords{dense matter-magnetic fields-methods: analytical-stars: individual(Quark Stars)-stars: magnetars}

\section{INTRODUCTION}
\label{sec:Intro}

The fundamental aspects of the physics involved in the description
of the matter inside a white dwarf are well understood
\citep{Shap_Teuk_book}, but in the case  of neutron stars the
situation is rather different because the equation of state (EoS)
of neutron matter at very high densities is still unknown.

The interior of a neutron star is an astrophysical laboratory in
which matter is compressed to high densities. The compression of
matter several times the saturation nuclear matter density,
$\rho_0$, may produce a phase transition from nuclear to quark
matter, i.e., an unconfined quark - gluon plasma. In addition,
under suitable circumstances, a conversion $d \to s$ quarks may
happen through weak interactions, leading to what has been called
strange quark matter (SQM). It has been stated that SQM may be the
absolute ground state of strong interactions \citep{bod,Witten},
although such hypothesis has not been confirmed yet. The natural
scenario where SQM could occur is the inner core of neutron stars.
Hence, if the SQM hypothesis is true, some neutron stars could be
either hybrid stars, which have quark cores surrounded by a
hadronic shell, or quark stars. Already 40 years ago, the
existence of quark stars in hydrostatic equilibrium was suggested
by \cite{Itoh} in a preliminary work.

On the other hand, it is well known that at the surface of neutron
stars there exist magnetic fields of the order of $10^{12} -
10^{13}\unit{G}$. Compact stars with ultra strong magnetic fields
($10^2-10^3$ larger than those of a typical neutron star) are
called magnetars. In such objects the magnetic field at the
surface could be higher than $10^{15} \unit{G}$ \citep{de la
Incera}.

The knowledge of the magnetars composition would help explain some
astrophysical phenomena. Soft gamma - ray repeaters (SGRs) and
anomalous X - ray pulsars (AXPs) have been interpreted as evidences
of magnetars. Besides some authors \citep{Cheng, Ouyed} claim
that magnetized hybrid or quark stars could be the real sources of
SGRs and AXPs. The M-R relationship tells us how matter composing
the star behaves under compression, providing information about
its composition. Several EoS for neutron, hybrid and quark stars
have been proposed but none of them is conclusive \citep{Haensel,
Latt01, Latt07, Ozel}. Each EoS produces a different mass-radius
(M-R) relationship which can be contrasted with the available
observational data in order to test its range of validity and/or
set bounds on some parameters. At this particular point
astrophysical studies become of great importance since they could
shed some light in understanding fundamental aspects of matter:
microphysics could be inferred from macrophysics. Here lies the
great importance of the studies related to ultra-compact objects. For
instance \cite{Latt01,Latt07} contrast some M-R relationships obtained
theoretically for different EoS. Varying some parameters a
difference of $4\%-10\%$ and $10\%-15\%$ in determining the maximum
radius $R_{max}$ and mass $M_{max}$, respectively, is shown for
the same EoS.

Several papers \citep{chakra,Daryel,aureola} study the M-R
relationship of highly magnetized quark stars (HMQS) through
numerical integration of the Tolman-Oppenheimer-Volkoff (TOV)
equation for different EoS.  Although most studies of quark stars
properties have used such method, \cite{bunjee} have obtained a
maximum mass and radius for unmagnetized quark stars analytically
by using a non-relativistic gravitational treatment.

Approximate analytical solutions play an important role in the
astrophysical analysis, giving a keener insight than the numerical
solutions. Moreover, they may be used as a testing point to check
if the numerical scheme is accurate and also they are the first step in the
comparison between theory and observation. Indeed, an approximate
analytical solution for M-R relationship may be all that is
required when comparisons with observational limits that determine
the confidence contour for the mass and radius are performed.
Besides, in the high - density EoS the uncertainties are of the same
order or larger than the errors in the variational method.

The appropriate treatment for quark stars should be relativistic,
since the existence of a maximum mass is associated with the
behavior of a relativistic gas and general relativistic
corrections are dominant \citep{weinberg_book}. In this paper, we
shall use the general relativistic energy variational principle
described by \cite{Chapline:1973} to obtain an analytic
approximate formula for the mass, radius, and baryonic number of an
HMQS. Quark stars are particularly suitable for a variational
treatment since their density profile resembles a constant mass
density star. We shall model an HMQS assuming quark matter within
high density regime in the framework of a modified MIT Bag model
EoS. We also assume that the magnetic field $B$ is low enough to
be treated like a correction in the EoS ($B<<\mu^2$, with $\mu$ being
the baryon chemical potential) although, as we will see in the
following sections, this is not a strong restriction.

The paper is organized as follow. In Section \ref{sec:QM:B} we
calculate the thermodynamical quantities of the system and
analyze the stability of quark matter with respect to
decomposition in baryons. In Section \ref{sec:MR} we provide the
analytic relativistic M-R relationship and compare our results
with the observational data. We also check the dynamic stability
of the star by calculating the adiabatic index and the speed of
sound . In Section \ref{sec:Summ:Conc}, we present a summary of our
main results and conclusions.

\section{High density quark matter within a strong
magnetic field} \label{sec:QM:B}

In this section we shall discuss the analytic approximations to
the SQM EoS in the presence of an uniform
magnetic field $\boldmath{B} \parallel \hat{\boldmath{z}}$. Within
the framework of the MIT Bag model, we assume three massless
quarks u, d, and s,  neglecting mediated interactions between them.
We also consider that the strong magnetic field is a small
contribution to the total energy, a fact that will be checked
later.

\subsection{Quark Matter in a Magnetic Field}
\label{ssec:QM:B}

Let us compute the grand canonical thermodynamic potential
$\Omega$ in the high density regime. Due the Landau quantization
 the phase space volume integral in the momentum space is replaced by
\begin{equation}
\frac{1}{(2\pi)^3}\int d\,^3p\,f(p)= \frac{1}{(2\pi)^3}\int dp_z\,
d\,^2p_\perp\, f(p)=\frac{qB}{4\pi^2}\,\sum_{\nu=0}^{\nu=\infty}
(2-\delta_{\nu 0})\int_{-\infty}^{+\infty} dp_z\, f(\nu,p_z),
\end{equation}
where $(2-\delta_{\nu 0})$ means that the zeroth Landau level is
singly degenerate, whereas all other states are doubly degenerate.
The grand canonical potential for each quark in the presence of a
strong magnetic field is given by
\begin{eqnarray}
\label{omega} \Omega_{i} &=& -\frac{q_i B g_i}{8\pi^2}
\sum_{\nu=0}^{\nu_{
    max}}(2-\delta_{\nu 0})\left[\mu \sqrt{\mu^2-2 \nu q_iB}-2
    \nu q_iB\ln\frac{\mu+\sqrt{\mu^2-2 \nu q_iB}}{\sqrt{2 \nu q_iB}}\right],
\end{eqnarray}
where $g_i=2 \times 3$, taking into account spin and color degeneracy, and $q_i$
is the absolute value of the charge of the particle, $q_{u} =
2|e|/3$ and $q_{d}=q_{s}= |e|/3$, with $e$ being the value of electronic
charge.

For simplicity, we consider the quark masses $m_q=0$, which
implies that the electrons are not present and quarks chemical
potential are, as a consequence of equilibrium conditions, all
equal, $\mu_u=\mu_d=\mu_s \equiv \mu$.

By imposing that
\begin{equation}
p_z^2=\mu^2-2 \nu q_iB \geq 0,
\end{equation}
we can determine the upper limit of the sum $\nu_{max}$ from
\begin{equation}
\nu \leq \frac{\mu^2}{2q_iB} \equiv \nu_{max}.
\end{equation}\label{numax}

The series, Equation (\ref{omega}), can be approximated using the
Euler-MacLaurin formula
\begin{eqnarray}
\label{maclau} \sum_{j=0}^{n}f(j)&=& \int_{0}^{n}f(x)
dx+\frac{1}{2}[f(n)+f(0)] \nonumber\\
& & + \frac{1}{12}[f'(n)-f'(0)]- \frac{1}{720}[f'''(n)-f'''(0)]+
R,
\end{eqnarray}
where the remainder term, $R$, usually is expressed in terms of
periodic Bernoulli polynomials \citep{math}, and can be estimated
by using
\begin{equation}
\left|R\right| \leq \frac{2 \zeta (4)}{(2 \pi)^4}
\int_1^{\nu_{max}-1} \left|f^{IV}(\nu)\right|\ d\nu,
\end{equation}
where the Riemann Zeta function $\zeta (4)\simeq 1.0823$. To avoid
divergences appearing in the third term of Equation
(\ref{maclau}),in the limit of high densities or
negligible quark masses, we apply the Euler-MacLaurin formula in the form
\begin{eqnarray}
\label{potexpand} \Omega_{i} &\simeq &
\Omega_i(\nu_{max})+\Omega_i(0)+\int_{1}^{\nu_{max}-1}\Omega_{i}(\nu)
d\nu +\frac{1}{2}\left[\Omega_i(\nu_{max}-1)+\Omega_{i}(1)\right] \nonumber\\
& & +\frac{1}{12}\left[\frac{\partial \Omega_i}{\partial \nu}\mid_{(\nu_{max}-1)}
  - \frac{\partial \Omega_i}{\partial \nu}\mid_{(1)}
  \right]+\tilde{R},
\end{eqnarray}
where
\begin{equation}
\label{reminnosso}
 \tilde{R}=-\frac{1}{720}\left[\frac{\partial^3
\Omega_i}{\partial \nu^3}\mid_{(\nu_{max}-1)}
  - \frac{\partial^3 \Omega_i}{\partial \nu^3} \mid_{(1)}
  \right]+R.
\end{equation}
We have considered Equation (\ref{reminnosso}) as the remainder
term, which gives $\mid \tilde{R}\mid \leq 3 \%$. In the limit
$\mu^2\gg 2 q_i B$, the thermodynamical potential can be calculated
performing first the integral in Equation (\ref{potexpand}) and
then expanding in power series of $B$. The result is

\begin{equation}
  \Omega _i = -{\frac {{\mu}^{4}}{{{4\pi}}^{2}}} +
  \,{\frac {{q_i}^2\,B^2}{{{8\pi}}^{2}}}\,\left(\mathrm{log}\frac{q_i\,B}{2\,\mu^2}-3\right)+ \mathcal{O}(B^{5/2}). \label{Omega_i}
\end{equation}
The particle density
$n_i=-\frac{\partial \Omega_{i}}{\partial\mu}$ is
\begin{equation}
n_i = {\frac {{\mu}^{3}}{{{\pi}}^{2}}}+\,{\frac
{{q_i}^2\,B^2}{{4\,{ \pi}}^{2}\,\mu}}+  \mathcal{O}(B^{5/2}). \label{n_i}
\end{equation}
Note that when $B=0$ in Equations (\ref{Omega_i}) and (\ref{n_i}), we
recover the usual expressions for a non-interacting massless quark
gas at zero temperature and zero magnetic field.

\subsection{Equation of State}
\label{ssec:EoS}

With the above results, one can form the modified EoS
of SQM in the MIT Bag model.  Within this framework, the
difference between the energy density of the perturbative and
non-perturbative QCD vacuum is taken into account by the ``bag
constant'' $\Bag$. Considering $\hbar = c = 1$ we can find the
conversion factor between high energy density units and magnetic
energy density units. We can write $\, 1 \, \unit{MeV} \,\simeq \,1.6 \,\times \,10^{-6}\,$ $\unit{erg} \,$ and
$\,1 \,\unit{MeV}\, \simeq \,(2 \,\times \,10^{-11}\, \unit{cm})^{-1}\,$, where $\,1 \,\unit{MeV}^4\, \simeq 2\, \times 10^{26}\,\, \unit{erg}\unit{cm}^{-3}$.
Relating this quantity with the magnetic energy density $\,B^2/8\pi\,$, the conversion factor for the
magnetic field is given by $\,1.4\, \times \,10^{13}\, \unit{G}\, \equiv \,1\,\, \unit{MeV}^2$.

The charge neutrality condition
\begin{equation}
\label{neutral} 2 n_u =  n_d + n_s
\end{equation}
and the $\beta$-equilibrium condition
\begin{equation}
  \label{eq:beta-eq}
  \mu_u=\mu_d=\mu_s \equiv \mu
\end{equation}
are automatically satisfied.

Combining the results of Section \ref{sec:QM:B} we obtain
\begin{equation}
\label{Pot_total}
\Omega=\sum_{i=u,d,s}\Omega_i+\mathrm{B}_{\mathrm{eff}}=-\frac{3
\mu^4}{4 \pi^{2}}+\frac
{B^2}{12{\pi}^{2}}\,\left(\mathrm{log}\frac{B}{2^{1/3}\,3\,\mu^2}-3\right)+\mathrm{B}_{\mathrm{eff}}+
\mathcal{O}(B^{5/2}),
\end{equation}
where $\mathrm{B}_{\mathrm{eff}}\,=\, \frac{B^2}{8 \pi}\,+\,\Bag\,$.
Replacing Equation (\ref{neutral}) in the baryon number density
condition, $n_B\,=\,\frac{1}{3}\,\sum_{i=u,d,s}\,n_i\,$, we obtain
\begin{equation}
\label{n_bar} n_B =\frac{\mu^3}{\pi^{2}}+\,{\frac
{B^2}{{9\,{ \pi}}^{2}\,\mu}}+  \mathcal{O}(B^{5/2}).
\end{equation}
Since we work in the $T\,=\,0$ limit, the energy density is given by
\begin{equation}
\label{energy}
 \rho = \Omega + 3 \mu n_B=\frac{9 \mu^4}{4
\pi^2}+\frac
{B^2}{12{\pi}^{2}}\,\left(1+\mathrm{log}\frac{B}{2^{1/3}\,3\,\mu^2}\right)+
\mathrm{B}_{\mathrm{eff}}+ \mathcal{O}(B^{5/2}),
\end{equation}
whereas the pressure reads
\begin{equation}
\label{pressure} P=-\Omega=\frac{3 \mu^4}{4 \pi^{2}}-\frac
{B^2}{12{\pi}^{2}}\,\left(\mathrm{log}\frac{B}{2^{1/3}\,3\,\mu^2}-3\right)-\mathrm{B}_{\mathrm{eff}}+
\mathcal{O}(B^{5/2}).
\end{equation}
Note that we are not considering the anisotropy of
pressures \citep{Orsa} because we are working in the limit of weak
magnetic field, $\mu^2 \gg 2 q_i
B$. The relation between the total energy density, Equation (\ref{energy}), and the total pressure, Equation (\ref{pressure}), determines the EoS of the system as
\begin{equation}
\label{rho} \rho = 3 P + 4 \mathrm{B}_{\mathrm{eff}}-\frac
{B^2}{3{\pi}^{2}}\,\left(2-\mathrm{Log}\frac{B}{2^{1/3}\,3\,\mu^2}\right)
+ \mathcal{O}(B^{5/2}).
\end{equation}

\subsection{Stability Analysis: Strong Interactions}

It is well known that SQM may be stable with respect to decay into
nucleons at zero pressure and zero temperature if its energy per
baryon $\frac{\rho}{n_B}$ is less than the energy per baryon of
$\null^{56}\mathrm{Fe}\, =\, 930 \,\unit{MeV}$ \citep{Farhi:1984}. The
presence of a magnetic field changes somewhat this stability
condition.

At $P\,=\,0$ we can estimate the chemical potential through successive approximation method as
\begin{equation}
\mu(B, \Bag)=\left[\frac{4 \pi^2
\mathrm{B}_{\mathrm{eff}}}{3}+\frac{B^2}{{3}^{2}}\left(\mathrm{log}\frac{B}{2^{4/3}\pi
\sqrt{3\Bag}}-3\right)\right]^{1/4},
\end{equation}
which will be replaced in equations (\ref{n_bar}, \ref{rho}) to
evaluate $\frac{\rho}{n_B}$. Contrary to previous results
\citep{Anand,chakra2,Daryel} we find that the energy per baryon
increases with $B$ (equation (\ref{rho})). The condition $\frac{\rho}{n_B}\,<\, 930\,\,MeV$ is satisfied for magnetic fields $B \,< \,4.4 \,\times \,10^{18}\,\, \unit{G}\,$. However for the stability of the system, not only it is necessary to consider the energy per baryon, but also the influence of magnetic energy density. We obtain $\,B^2/8\pi\, \sim\, \Bag\,$ for $\,85\,\,\unit{MeV}\unit{fm}^{-3}\,<\,\Bag\,<\,90\,\,\unit{MeV}\unit{fm}^{-3}$ and $\,B^2/8\pi\, > \,\Bag\,$ for $\,57\,\,\unit{MeV}\unit{fm}^{-3}\,<\,\Bag\,<\,80\,\,\unit{MeV}\unit{fm}^{-3}$. Although it is known that the binding
of the quark stars is provided not by gravitation,
but rather by the strong interactions, the inclusion of a
magnetic field $B$ adds an additional constraint to the stability condition through the magnetic energy density and the magnetic pressure. When the latter is not
lower than $\Bag$ the magnetic field becomes dynamically important. Furthermore if the magnetic pressure is of the same order
of magnitude than the matter pressure, spherical deformation effects
should be considered. In addition sufficiently strong magnetic fields can
generate an anisotropic pressure
distribution inside the HMQS modifying the EOS and consequently
the M-R relationship \citep{Paulucci}.
Therefore for magnetic fields
large enough, $\mathrm{B}_{\mathrm{eff}}$ will be greater than
the kinetic energy of the quarks thereby destabilizing the star.
Thus we also consider the relationship
\begin{equation}
\label{ss} (B^2/8\pi)\,\Bag^{-1} < 0.1
\end{equation}
to guarantee both the perturbative treatment of the system and the
stability of the star. Table \ref{table1} shows that the variation
of baryon density is quite small for $P\,=\,0$ when increasing the
magnetic field from 0 up to $B_{max}$.

\section{MASS-RADIUS RELATIONSHIP BY VARIATIONAL METHOD}
\label{sec:MR}

The energy variational method in general relativity is explained
in detail in (\cite{Harrison.el.al:1965}; see also \cite{weinberg_book}). Starting from an
uniform density configuration in a spherically symmetric
distribution the total mass $M$, the baryon number $N_B$ and the
radius $R$ of the star are given by
\begin{eqnarray}
\label{Mst} M &=& \frac{4}{3} \pi \rho R^3 ,  \nonumber \\ N_B &=& 2\pi n_B
a^{3} (\chi - \sin \chi
\cos \chi),  \\
R &=& a \sin \chi \, \nonumber
\label{Rst}
\end{eqnarray}
where $\rho$ is the mass-energy density and the angle $\chi$ comes
from substituting $r\, =\, a \,\sin\,\chi$, where $a\, = \,[\,3/(8 \pi
\rho)\,]^{1/2}$ is the curvature radius in the metric inside the
star which adopts the following form for the 3-geometry:
\begin{equation}
ds^2 = a^2 \left[d\chi^2+\sin ^2 \chi \left( d \theta ^{\,2} + \sin ^2
\theta d \phi ^2\right) \right].
\end{equation}
Note that we are using $\hbar = c = G = 1$. The configuration of
maximum density is achieved when $\chi=\pi/2$. Observe that $\sin
^2 \chi \,=\, 2\,M/R\,$, $\chi \,\sim \,0\,$ corresponds to the Newtonian limit
while $\chi = \pi/2$ corresponds to the Schwarzschild one. The use
of a constant energy trial configuration has been justified by
\cite{Chapline:1973}, while \cite{weinberg_book}has applied it to
white dwarfs. In this latter case, a fair approximation to the
Chandrasekhar mass was obtained.

\begin{figure}
\epsscale{0.80} \plotone{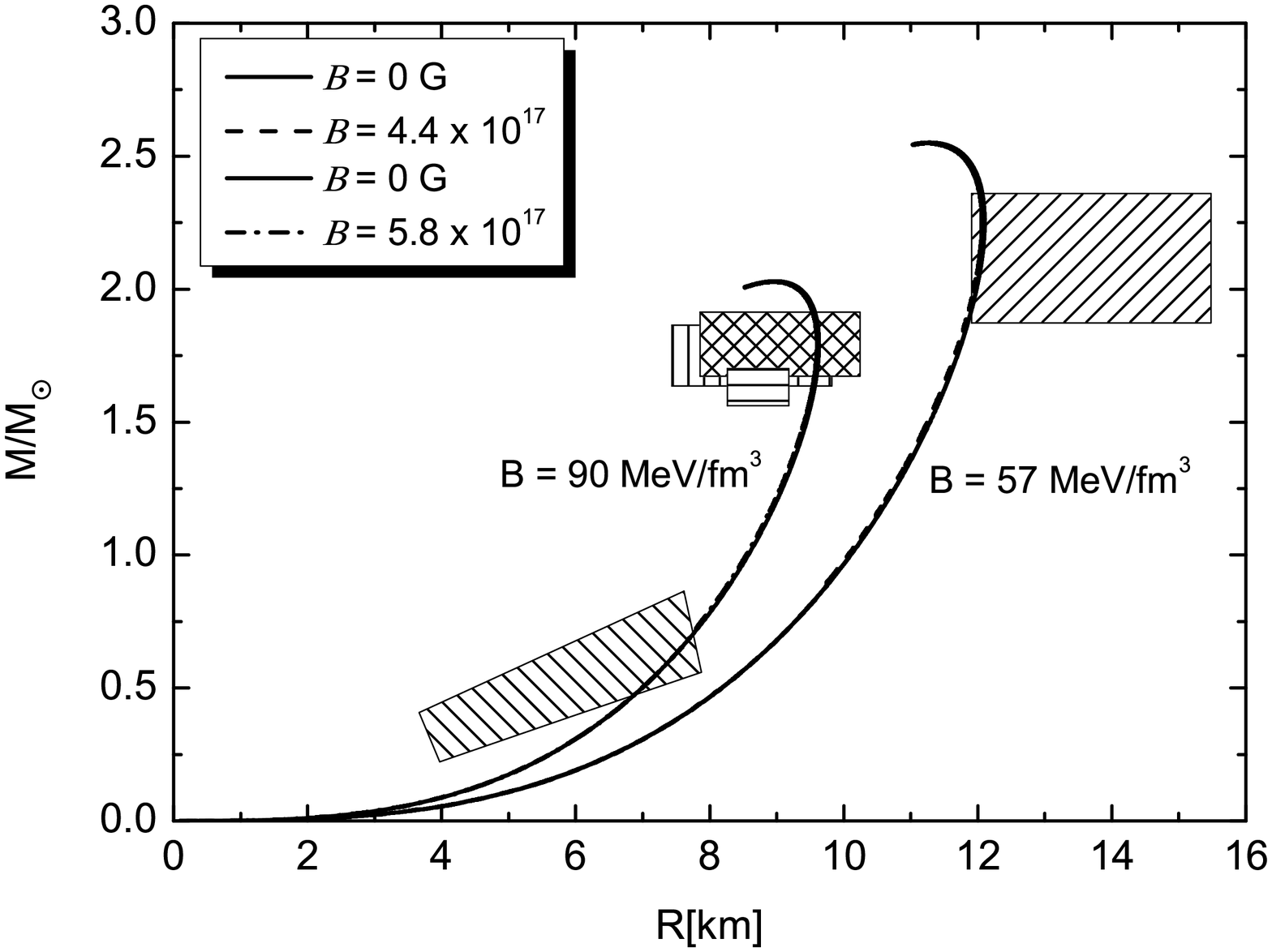} \caption{Mass-Radius
relationship with and without magnetic field for $\Bag=57
\unit{MeV}\unit{fm}^{-3}$ (solid and dashed line) and $\Bag=90
\unit{MeV}\unit{fm}^{-3}$ (solid and dash-dotted line). The
rectangle with diagonal pattern corresponds to EXO 0748-676, which has been
interpreted as a hadronic star. Rectangles with crossed, vertical,
and horizontal patterns correspond to quarks or hybrid stars
\citep{drago}. The polygon could be a low-mass strange star as
suggested in \citep{Zhang}.\label{MasaRad}}
\end{figure}

To obtain the equilibrium condition is appropriate to treat $\chi$
as an independent variable. Imposing $\partial\, M\,/\,\partial \,\chi\,=\,0\,$
for fixed $N_B$, the equilibrium condition reads
\begin{equation}
\label{zetaeqw} w \equiv \frac{P}{\rho} = \zeta (\chi ),
\end{equation}
where $\rho$ and $P$ are given by Equations (\ref{energy}) and
(\ref{pressure}), and $\zeta (\chi )$ is a function independent of
the EoS
\begin{equation}
\zeta (\chi) = 3\,\cos \chi   \left( \frac{9}{2}\,\cos  \chi  -{
\frac { \sin ^3 \chi}{\chi-\sin
  \chi \cos \chi }} \right) ^{-1}-1.
\end{equation}
We get an approximate value of $\,\zeta (\chi)\,$ using a Taylor
series, $\,\zeta _{T}\,$, around $\,\chi\,=\,0\,$. Truncation at eighth order gives
\begin{eqnarray}
\label{taylor} \zeta _{T} &=& \frac{1}{10}\,{\chi}^{2}+{\frac
{113}{2100}}\,{\chi}^{4}+{\frac {1747}{63,000}} \,{\chi}^{6}+{\frac
{689,687}{48,510,000}}\,{\chi}^{8}.
\end{eqnarray}
The Pad{\'e} approximant of order $(4\,,\,4) $ gives a representation
of this function that is also an approximate analytic continuation
beyond the circle of convergence. Thus, $\zeta (\chi)$ is given as
a ratio of two polynomials as
\begin{eqnarray}
\label{pade} \zeta _{P} &=& \left( -{\frac
{23}{6237}}\,{\chi}^{4}+\frac{1}{10}\,{\chi}^{2} \right)
 \left( 1-{\frac {5123}{8910}}\,{\chi}^{2}+{\frac {3002}{93,555}}\,{
\chi}^{4} \right) ^{-1}.
\end{eqnarray}
Imposing $\zeta _{P}=w$ we obtain the only physical solution for
$\chi$, always positive and fulfilling the condition $\,\lim_{w \,\to\,
0}\, \chi \,=\, 0\,$,  given by
\begin{equation}
\label{chi} \chi = \frac{\sqrt {3}}{2}\,{\frac {\sqrt { \left(
35861\,w+6237-\sqrt {786718681\,w^{2}+389949714\,w+38900169}
 \right) }}{\sqrt{3002\,w+345}}}.
\end{equation}
Hence we get an analytical expression of the HMQS mass and radius
as a function of the baryonic chemical potential. This allows us
to obtain the M-R relationship for different Bag constant and
magnetic field values. In particular in Figure \ref{MasaRad} we
show the $\,\Bag\, = \,57\,\, \unit{MeV}\unit{fm}^{-3}$ and $\,\Bag \,= \,90\,\,
\unit{MeV}\unit{fm}^{-3}$ cases with and without magnetic field.
Note that for the first one, when $B\,=\,0$, the M-R curve coincides
with the hadronic star zone. This result can be attributed to the
fact that the value $\,\Bag\, = \,57 \,\,\unit{MeV}\unit{fm}^{-3}$ could be too low
for a quark star which has been modeled by using the MIT bag
model \citep{Zdunik2000}. Furthermore, in Table \ref{tab2} we
present the results for other values of $\Bag$. Note that although
$B_{max}$ is a typical value for a magnetar it slightly decreases
the $M_{max}$ if compared with the zero magnetic field case. The same result is obtained for $R_{max}$.

\subsection{Dynamical Stability}
In our model the condition for stable equilibrium is given by
$\partial^2\, M\,/\,\partial^2\, \chi\, > \,0$. For a given EoS, it is
possible to determine the quark densities and pressures where quark
stars are stable against gravitational collapse from the condition
\begin{displaymath}
\Gamma>\Gamma_c,
\end{displaymath}
where the adiabatic index for SQM, $\Gamma$, is given by:
\begin{eqnarray}
\Gamma = \frac{n_B}{P}\frac{dP}{dn_B}&=&\frac{4 \mu^4}{3
\mu^4-4\pi^2\mathrm{B}_{\mathrm{eff}}}
-\frac{2B^2\mu^4}{9}\,\,\frac{(7-6\,\mathrm{Log}\frac{B}{2^{1/3}\,3\,\mu^2})}{(3
\mu^4-4\pi^2
\mathrm{B}_{\mathrm{eff}})^2}\nonumber\\
& & -\frac{88\,B^2}{27}\,\,
\frac{\mathrm{B}_{\mathrm{eff}}\,\pi^2}{(3 \mu^4-4\pi^2
\mathrm{B}_{\mathrm{eff}})^2}+ \mathcal{O}(B^4) ,
\end{eqnarray}
and the critical adiabatic index, $\Gamma_c$, for a cold star in
general relativity is
\begin{equation}
  \Gamma_c=(1+w)\left[1+\frac{(3 w+1)}{2}\left[\frac{(w+1)}{6
        w}\tan^2\chi-1 \right] \right].
\end{equation}
To get dynamical stability the condition $\Gamma>\frac{4}{3}$ must
be satisfied. The intersection between $\Gamma$ and $\Gamma_c$
determines when quark star becomes gravitationally unstable. We
found $\Gamma_c \sim 2.3$ and $w_c\sim 0.17$ for the maximum mass
value of the star (Table \ref{tab2}).

Another quantity that is related with the stability of the star is
the speed of sound $c_s$. To satisfy the causality of quark matter,
\begin{equation}
\frac{dP}{d\rho}=c_s^2 \leq 1.
\end{equation}
At very high densities particles become relativistic and the speed of sound should be lower,
 more precisely of the order of $1/\sqrt{3}$, the speed of sound of relativistic fluids. We found
\begin{equation}
\label{sound}
c_s=\frac{1}{\sqrt{3}}+\frac{B^2}{27\sqrt{3} \mu^4} +
\mathcal{O}(B^2).
\end{equation}
This quantity tell us how stiff is the EoS providing information
about the compressibility of the fluid. The stiffness of the EoS increases when $c_s$ is closer to 1.

\section{SUMMARY AND CONCLUSIONS}
\label{sec:Summ:Conc}

In this article we have furnished an analytical treatment to study
an HMQS in the framework of the MIT Bag model. We have analyzed the
stability of quark matter with respect to strong interactions and
 found a restriction in the stability condition: there is a
maximum value for the magnetic field, $B_{max}$, beyond which
quark matter becomes unstable. In the limit of ``weak'' magnetic
field that we have studied, quark magnetic moments are aligned in
the same direction of the field and this situation leads to such
restriction. This result could mean that if the magnetic field
strength exceeds that critical value, then quark or hybrid stars
should not be considered as magnetars. In a more general case,
when quark masses are take into account, electrons should be
considered. For magnetic fields much stronger than $B_{max}$,
Landau levels for electrons will increase the energy per particle.
However this probably will not contribute to modify the
iron-condition , $\null^{56}\mathrm{Fe}\, =\, 930 \,\unit{MeV}$,
because electron fraction in SQM is already very low.

We have also found an analytical approximate solution for the M-R
relationship. Even tough we used very simple physics, our results are in
good agreement with the confidence contours of available
observational data.

It is important to note that although the uniform energy density regime is a
good approximation for quark stars, deviations in the
determination of M-R relationship may occur because in the
limit of high densities such approximation is no longer valid.

Finally, we calculate the adiabatic index and the speed of sound.
The critical value for the adiabatic index, which corresponds to
the collapse of the star, is in agreement with that of
\cite{Chapline:1977}, a pioneering work about quark stars. On the
other hand, the speed of sound  is consistent with the expected
values for quark stars.

\acknowledgments

We thank A. P\'erez Martínez and J. E. Horvath for comments and suggestions.
M.O. acknowledges the fruitful discussion with F. Weber and H.
Rodrigues. We are grateful to A. Reisseneger for kindly and
carefully answering our questions on the high energy units.
We also thank S. Chakrabarty.

\clearpage

\clearpage

\begin{table}[h]
\caption{Bag Constant, Baryon Density and Magnetic Field Upper Limit to Preserve Quark Matter Stability Condition.
\label{table1}}
\begin{center}
\begin{tabular}{ccc}
\tableline\tableline
 $\Bag$[MeV$fm^{-3}$] & \hspace*{.1cm}$n_B/n_0$ \hspace*{.1cm} &$B_{max}$ [G]\\
\tableline
57 & $1.73 \pm 0.01$ & $4.4 \times 10^{17}$ \\
60 & $1.80 \pm 0.01$ & $4.8 \times 10^{17}$ \\
75 & $2.14 \pm 0.01$ & $5.3 \times 10^{17}$ \\
80 & $2.24 \pm 0.02$ & $5.5 \times 10^{17}$\\
85 & $2.34 \pm 0.02$ & $5.7 \times 10^{17}$ \\
90 & $2.45 \pm 0.02$ & $5.8 \times 10^{17}$ \\
\tableline\tableline
\end{tabular}
\end{center}
\end{table}

\clearpage

\begin{table}[h]
\caption{Maximum Mass, Maximum Radius and Baryonic Number for Different
Bag Constants. \label{tab2}}
\begin{center}%
\begin{tabular}[c]{ccccccc}
\tableline \tableline

                      \hspace*{.1cm}  $\Bag$ $[MeVfm^{-3}]$ \hspace*{.1cm}
   & \hspace*{.3cm}  $B$ $[G]$ \hspace*{.3cm}  & \hspace*{.2cm}  $R_{max}$ $[km]$ \hspace*{.2cm}  &   \hspace*{.2cm}  $M_{max}/M_{\odot}$  \hspace*{.2cm} &   \hspace*{.2cm}  $N_B/N_{\odot}$  \hspace*{.2cm}   \\ \hline
 \multirow{7}{.3cm}[15mm]{57}                       &  0  & 12.10    &    2.55     &     4.30      & \\ \cline{2-6}
       &  4.4 $\times$ $10^{17}$  & 12.06   &    2.45    &     4.09     & \\ \cline{1-6}

\hline
  \multirow{7}{.3cm}[15mm]{60}                    &  0  &  11.80    &    2.49     &     4.14       &  \\  \cline{2-6}
                 &   4.8 $\times$ $10^{17}$ &  11.76 &    2.48     &     4.14       & \\ \cline{1-6}
\hline
 \multirow{7}{.3cm}[15mm]{75}                     &  0  & 10.55   &    2.22    &     3.50       &  \\  \cline{2-6}
                 &   5.3 $\times$ $10^{17}$ &  10.52 &    2.22     &     3.50      & \\ \cline{1-6}
 \hline
  \multirow{7}{.3cm}[15mm]{80}                    &  0  &  10.22   &    2.15    &     3.34       &  \\  \cline{2-6}
                 &   5.5 $\times$ $10^{17}$ &  10.18 &    2.15     &     3.34       & \\ \cline{1-6}
\hline
     \multirow{7}{.3cm}[15mm]{85}                 &  0  &  9.91    &    2.09    &     3.19       &  \\  \cline{2-6}
                 &   5.7 $\times$ $10^{17}$ &  9.89 &    2.09    &     3.19       & \\ \cline{1-6}
\hline
    \multirow{7}{.3cm}[15mm]{90}                  &  0  &  9.63   &    2.03   &     3.05      &  \\  \cline{2-6}
                 &   5.8 $\times$ $10^{17}$ &  9.60 &    2.03     &     3.05       & \\ \cline{1-6}

\tableline\tableline
\end{tabular}
\end{center}
\end{table}



\end{document}